\documentclass[12pt]{article}
\usepackage{graphicx,subfigure,psfrag}
\usepackage{amsfonts,amsmath}

\usepackage{color}
\usepackage{tabularx}
\newcommand{\pc}[1]{\parbox{0pt}{#1}}
\usepackage[colorlinks=false]{hyperref} 
\usepackage{dsfont}
\usepackage{times}

\newcommand{\complex}{{\mathds C}} 

\def\+{{+\!\!\!+}}

\def\P{\Phi}

\newcommand{\C}{{\mathds C}}

\def\F{{\cal F}}
\def\T{{\rm T}}

\def\Tr{\rm Tr}
\def\log{\rm log}

\def\P{{\cal P}}

\def\pmb#1{\setbox0=\hbox{#1}%
\kern.0em\copy0\kern-\wd0 
\kern-.04em\copy0\kern-\wd0 
\kern.08em\copy0\kern-\wd0 
\kern-.04em\raise.0433em\box0 }         
 
\def\diag{\textstyle{\rm{diag}}} 
\newcommand{\nc}{\newcommand} 
\nc{\beq}{\begin{equation}} 
\nc{\eeq}[1]{\label{#1}\end{equation}} 
\nc{\ber}{\begin{eqnarray}} 
\nc{\eer}[1]{\label{#1}\end{eqnarray}} 
\nc{\pek}[1]{\cite{#1}} 
\nc{\enr}[1]{(\ref{#1})} 
\nc{\kal}[1]{{\cal{#1}}} 
\nc{\dott}{\;\cdot\;} 

\def\0 {\nonumber}

\begin{document} 

\setcounter{page}{0}
\newcommand{\inv}[1]{{#1}^{-1}} 
\renewcommand{\theequation}{\thesection.\arabic{equation}} 
\newcommand{\be}{\begin{equation}} 
\newcommand{\ee}{\end{equation}} 
\newcommand{\bea}{\begin{eqnarray}} 
\newcommand{\eea}{\end{eqnarray}} 
\newcommand{\re}[1]{(\ref{#1})} 
\newcommand{\qv}{\quad ,} 
\newcommand{\qp}{\quad .} 

\def\qp{Q_+}
\def\qm{Q_-}
\def\qbp{\bar Q_+}
\def\qbm{\bar Q_-}
\def\sgh{\Sigma_{g,h}}

\begin{titlepage} 
\begin{center} 

\hfill SISSA 26/2009/FM-EP\\  
                         
\vskip .3in \noindent 


{\Large \bf{Decoupling A and B model in open string theory}}


{\large {\it Topological adventures in the world of tadpoles}} 

\vskip .3in 

{\bf Giulio Bonelli, Andrea Prudenziati, Alessandro Tanzini, and Jie Yang}

\vskip .05in 
{\em\small International School of Advanced Studies (SISSA) \\ and \\ INFN, Sezione di Trieste \\
 via Beirut 2-4, 34014 Trieste, Italy} 
\vskip .5in
\end{center} 
\begin{center} {\bf ABSTRACT }  
\end{center} 
\begin{quotation}\noindent  
In this paper we analyze the problem of tadpole cancellation in open topological strings.
We prove that the inclusion of unorientable worldsheet diagrams guarantees
a consistent decoupling of A and B model for open superstring amplitudes at all genera.
This is proven by direct microscopic computation in Super Conformal Field Theory. 
For the B-model we explicitly calculate one loop amplitudes in terms of analytic Ray-Singer torsions of appropriate vector 
bundles and obtain that the decoupling corresponds to the cancellation of D-brane and orientifold charges.
Local tadpole cancellation on the worldsheet then guarantees the decoupling at all loops. 
The holomorphic anomaly equations for open topological strings at one loop are also obtained and compared 
with the results of the Quillen formula.
\end{quotation} 
\vfill 
\eject

\end{titlepage}

\tableofcontents

\section{Introduction}

It is a classical result in open string theories that 
the condition of tadpole cancellation ensures their consistency
by implementing the cancellation of gravitational and mixed anomalies
\cite{Polchinski_I}.
It is also well known that topological string amplitudes calculate
BPS protected sectors of superstring theory
\cite{bcov,AGNT,ant}.
It is therefore natural to look
for a corresponding consistency statement in the open topological string.

Closed topological strings on Calabi-Yau threefolds provide a beautiful description 
of the K\"ahler and complex moduli space geometry via the A- and B-model respectively
\cite{Witten:1991zz}.
However, D-branes naturally couple in these models to the wrong moduli \cite{OOY}, namely
A-branes to complex and B-branes to K\"ahler moduli.
This leads to new anomalies in the topological string due to boundary terms, as observed
in \cite{jie} and constitutes 
an obstruction to mirror symmetry and to the realization of open/closed string duality in generic Calabi-Yau targets.
A proper analysis of this problem is thus compelling. 
In this paper we will show how to cancel these new
anomalies at all loops by including crosscap states. 
We will also find that from the target space
viewpoint, this corresponds to the cancellation of
D-brane and orientifold charges.

The first observation in this direction came from a different perspective in \cite{evidence}, where it
was observed that the inclusion of unorientable
worldsheet contributions is crucial to obtain a consistent BPS
states counting for some specific geometries in the open A model \cite{jw,kw1}.
From this it was inferred that tadpole cancellation would ensure the decoupling of A and B model
in loop amplitudes.
In this paper we provide a Super Conformal Field Theory derivation of the above statements. 

We also provide a target space geometric interpretation for unorientable one loop amplitudes
in open B-model in terms of analytic Ray-Singer torsions \cite{RS}.
This allows us to show explicitly that the decoupling of K\"ahler moduli corresponds to the cancellation 
of D-brane and orientifold charge.

Some analysis on the unoriented sector of the topological string have been performed in
\cite{SV,bobby,Diaco,Bou} for local Calabi-Yau geometries. In these cases the issue of tadpole cancellation gets easily solved by 
adding anti-branes at infinity, as already noticed also in \cite{jie}.
However, a more systematic study of this problem is relevant in order to analyze mirror symmetry with D-branes \cite{vafa} 
and open/closed string dualities in full generality.

More in general, it is expected that the topological string captures
D-brane instanton non perturbative terms upon Calabi-Yau compactifications to four dimensions \cite{Ma,u}.
Therefore, the study of the geometrical constraints following from a consistent wrong moduli decoupling could shed light 
on the properties of BPS amplitudes upon wall crossing \cite{KS,JM,OY,NN,GM}.

The structure of the paper is the following.
In section \ref{tad1} we consider tadpole cancellation at one loop 
in the simple case of target $T^2$ and 
rewrite the resulting
amplitudes in terms of Ray-Singer torsions of suitable vector bundles. 
In section \ref{hae} we move to a generic Calabi-Yau target space by considering
the complete set of holomorphic anomaly equations and discussing 
tadpole cancellation at one loop.
In section \ref{storsion} 
we continue the microscopic analysis by directly calculating  unoriented one loop B-model 
amplitudes on a generic Calabi-Yau threefold in terms of 
analytic Ray-Singer torsions of appropriate vector bundles. We show that
the requirement of decoupling of wrong moduli corresponds to tadpole
cancellation.
In section \ref{all} we extend our arguments to all loops and show how local tadpole cancellation on the world-sheet
absorbs the disk function anomaly observed in \cite{jie}.
We leave our concluding observations for section \ref{conclusion}.

\section{One loop amplitudes on the torus, tadpole cancellation and Ray-Singer analytic torsion}
\label{tad1}

In this section we investigate tadpole cancellation
for open unoriented topological string amplitudes at zero Euler characteristic
considering, as a warm up example, the B-model case when the target space is a $T^2$.
We conclude by rewriting the amplitudes as Ray-Singer analytic torsions.

The relevant amplitudes are 
the cylinder, the M\"obius strip and the Klein bottle coupled to a constant gauge field. 
In the operator formalism, as usual for one loop amplitudes, we have
\begin{equation*}
\F_{cyl} = \int_{0}^{\infty}\frac{ds}{4s}{\Tr}_{o}\left( F(-1)^{F}e^{-2\pi s H}\right),\:\:\:\: 
{\cal F}_{m\ddot{o}b} = \int_{0}^{\infty}\frac{ds}{4s}{\Tr}_{o}\left(\P F(-1)^{F}e^{-2\pi s H}\right) 
\end{equation*}
\begin{equation} \label{d}
\F_{kle} = \int_{0}^{\infty}\frac{ds}{4s}{\Tr}_{c}\left(\P F(-1)^Fe^{-2\pi s H}\right) 
\end{equation}
where $\P=\Omega\circ\sigma$ is the involution operator obtained by combining the worldsheet parity operator $\Omega$
and a target space involution $\sigma$, $F$ is the fermion number and $H$ is the Hamiltonian for worldsheet time translations. 
The trace is taken over all ( open or closed ) string states. In this section we consider D-branes wrapping the whole $T^2$
and take $\sigma$ to act trivially.
From the Hamiltonian $H$ of the $\sigma$-model with Wilson lines for gauge groups $SO(N)$ or $Sp(N/2)$, 
we have (setting $\alpha' = 1$):
\begin{equation} \label{a}
\F_{cyl} = \sum_{n,m = -\infty}^{+\infty}\:\sum_{i,j = 1}^{N}\int_{0}^{\infty}\frac{ds}{4s}\: e^{-\frac{2\pi s}{\sigma_{2}t_{2}}\left|n -\, \sigma m - u_{i,j} \right|^2}
\end{equation}
\begin{equation} \label{b}
\F_{m\ddot{o}b} = \pm \sum_{n,m = -\infty}^{+\infty}\:\sum_{i = 1}^{N}\int_{0}^{\infty}\frac{ds}{4s}\: e^{-\frac{2\pi s}{\sigma_{2}t_{2}}\left|n -\, \sigma m - 2u_{i} \right|^2}
\end{equation}
\begin{equation} \label{c}
\F_{kle} = \sum_{n,m = -\infty}^{+\infty}\int_{0}^{\infty}\frac{ds}{4s}\: e^{-\frac{2\pi s}{2\sigma_{2}t_{2}}\left|n - \,\sigma m \right|^2}
\end{equation}
 
Here $u_{i,j} = u_{i} - u_{j}$ and $u_{i} = \phi_{i} + \sigma\theta_{i}$ with $\theta_{i}$ and $-\phi_{i}$ the $i$-th diagonal element 
of the Wilson lines 
\footnote{As reviewed in the Appendix the unoriented theory selects either the $Sp(N/2)$ or the $SO(N)$ groups. 
In both cases one can diagonalize with a constant gauge transformation leading to N diagonal elements. 
These are purely imaginary for $SO(N)$ and real for $Sp(N/2)$, half of them being independent numbers $a_{1},...,a_{N/2}$ 
and the other half $-a_{1},...,-a_{N/2}$.} 
along the two 1-cycles of the torus with complex structure 
$\sigma = \sigma_{1} + i \sigma_{2} = \frac{R_{2}e^{i\rho}}{R_{1}}$ 
and area $t_{2} = R_{1}R_{2}\sin(\rho)$. 
This means that if one parametrizes the target space torus with $z = R_{1}x_{1} + R_{2}e^{i\rho}x_{2}$, then the gauge field 
reads $A_{i} = \theta_{i}dx_{1} - \phi_{i}dx_{2}$. 
The topological amplitudes get contribution from classical momenta only, due to 
a complete cancellation between the quantum bosonic and fermionic traces. 
The shift in the classical momenta by the Wilson lines $u_{i}$ is the only effect of the coupling to the gauge fields. 
Note that the different coupling between the cylinder and the M\"obius is due to the selection of diagonal $\P$ 
states for the M\"obius. The $\pm$ in front of the M\"obius corresponds to the $SO(N)$ and $Sp(N/2)$ theories respectively
coming from the eigenvalues of the Chan-Paton states in the trace under $\P$.

These amplitudes suffer of two kinds of divergences: the first one is from the $ s\rightarrow 0 $ part of the integral 
and will be removed by tadpole cancellation. The second one comes from the series which turns out to diverge for
vanishing Wilson lines \cite{RS}. In the superstring this second divergence is due to extra massless modes generated by gauge 
symmetry enhancement.
We will start with tadpole cancellation and deal later with the second divergence.

In order to analyze
the behaviour at $ s\rightarrow 0 $, we 
Poisson resum the $n,m$ sums in order to get an exponential going like $e^{-1/s}$. The result is:
\begin{equation} \label{e}
\F_{cyl} = \sum_{m,n = -\infty}^{+\infty}\:\sum_{i,j = 1}^{N}\int_{0}^{\infty}ds\frac{t_{2}}{8s^2}\: 
e^{-\frac{\pi t_{2}}{2s\sigma_{2}}\left|n + \sigma m\right|^2}e^{2\pi i\left( m\phi_{i,j} - \,n\theta_{i,j}\right)}
\end{equation}
\begin{equation} \label{f}
\F_{m\ddot{o}b} = \pm\sum_{m,n = -\infty}^{+\infty}\:\sum_{i = 1}^{N}\int_{0}^{\infty}ds\frac{t_{2}}{8s^2}\: 
e^{-\frac{\pi t_{2}}{2s\sigma_{2}}\left|n + \sigma m\right|^2}e^{2\pi i\left( 2m\phi_{i} - \,2n\theta_{i}\right)}
\end{equation}
\begin{equation} \label{g}
\F_{kle} = \sum_{m,n = -\infty}^{+\infty}\:\int_{0}^{\infty}ds\frac{t_{2}}{4s^2}\: 
e^{-\frac{\pi t_{2}}{s\sigma_{2}}\left|n + \sigma m\right|^2}
\end{equation}
In order to extract the tadpole divergent part, let us
perform the change of variables $\frac{\pi}{s} \rightarrow s$,  $\frac{\pi}{4s} \rightarrow s$, and $\frac{\pi}{2s} \rightarrow s$
respectively for cylinder, M\"obius and Klein bottle 
\footnote{This is in order to normalize the three surfaces to have the same circumference and length 
(respectively $2\pi$ and $s$). They are
parametrized such that the M\"obius and the Klein are cylinders with one and two boundaries substituted by crosscaps respectively.}.
In the three cases the divergent parts come from the $n = m = 0$ term and adding the three contributions we get
\[
\int_{0}^{\infty}ds\frac{t_{2}}{8\pi}\:\left(N^2 \:\:( cylinder )\:\:\ \pm 4N \:\: ( M\ddot{o}bius ) \:\: + 4 \:\:( Klein )\right).
\]
The divergence is canceled by choosing $N = 2$ and requiring $Sp(N/2)$ gauge group
\footnote{In the case of target space $T^{2d}$ one finds $N=2^d$.}.

Once this divergence is removed, the M\"obius strip with non-zero Wilson lines is finite and 
reads
\begin{equation}
{\cal F}_{m\ddot{o}b} = +\frac{1}{2}log\prod_{i = 1}^{N}\left|e^{\pi i (2\theta_{i})^{2}\sigma}\theta_{1}
\left(2u_{i}|\sigma \right)\eta\left( \sigma\right)^{-1}\right|.
\label{moebtorus}
\end{equation}
in terms of the standard modular functions $\theta_1$ and $\eta$.

As it is evident from (\ref{moebtorus}), a further divergence arises at vanishing Wilson lines, where $\theta_1$ vanishes.
In order to define a finite amplitude, notice that
for small value of one of the $u_{i}$'s we can expand to first order inside the logarithm getting
\bea
\F_{m\ddot{o}b}^{reg} = ... + \frac{1}{2}log\left|e^{\pi i (2\theta_{i})^{2}\sigma}
\theta_{1}\left(2u_{i}|\sigma \right)\eta\left( \sigma\right)^{-1}\right| + ... \approx
\\
\approx ... + \frac{1}{2}log\left| 0 - 2\pi i \eta
\left( \sigma\right)^{2}\sqrt{\sigma_{2}}\frac{2u_{i}}{\sqrt{\sigma_{2}}} + ... \right| + ... 
\label{cacca}\eea
Notice that both $\eta\left(\sigma\right)^{2}\sqrt{\sigma_{2}}$ and $\frac{2u_{i}}{\sqrt{\sigma_{2}}}$ are separately 
modular invariant under the $SL(2,{\mathbb Z})$ transformations
\footnote{Recall that $u_{i} = \phi_{i} + \sigma\theta_{i}$ and $- \phi$ and $\theta$ are the gauge fields along the two cycles.}
:
\[
\sigma \rightarrow \frac{c + d\sigma}{a + b\sigma} \:\:\:\:\:\:\:
\theta \rightarrow a\theta - b \phi \:\:\:\:\:\:\:
\phi \rightarrow -c\theta +d\phi 
\]
From (\ref{cacca}), it is clear that the remaining 
finite part is the $\eta\left(\sigma\right)^{2}\sqrt{\sigma_{2}}$ term in the logarithm. One can
in fact compute it for vanishing Wilson lines starting from (\ref{b}) by first regulating the integral as
\begin{equation} \label{h}
\int_{\epsilon}^{\infty}\frac{dt}{t}e^{-kt} = C -log(k\epsilon) + O(\epsilon) \approx -log(k) + C
\end{equation}
and discarding the $m = n = 0$ term, which take care of the tadpole divergence. Then by using zeta-function regularization 
to deal with the infinite product over the $k$-factors in the logarithm one gets\footnote{We use the formula 
$\sin(\pi z) = \pi z \prod_{n = 1}^{\infty}\left(1 - \frac{z^{2}}{n^{2}} \right)$.}
\begin{eqnarray}
\F_{m\ddot{o}b} =  +\frac{1}{4}log\prod_{i= 1}^{N}\frac{\sigma_{2}t_{2}}{2\pi 4 u_{i}\overline{u}_{i}}
\left| e^{i\pi 2u_{i} } - e^{-i \pi 2u_{i}} \right|^{2}e^{2 \pi i \sigma /12}e^{-2 \pi i \overline{\sigma} /12}\times\nonumber
\\
\times\prod_{m = 1 }^{\infty}
\left| \left(1 - e^{2 \pi i (m\sigma + 2u_{i})}\right)\left(1 - e^{2 \pi i (m\sigma - 2u_{i})}\right)\right|^{2}+C.
\label{zio}
\end{eqnarray}
This is well behaved for $u_{i} \rightarrow 0$ 
giving, for each vanishing Wilson line element, a term
\begin{equation} \label{i}
 \frac{1}{2}log\left(\frac{\sqrt{\sigma_{2} t_{2}}}{\sqrt{2}}\left|\eta(\sigma)\right|^{2}\right) +\frac{1}{4}log(4\pi) + C.
\end{equation}
The constant $C$ is arbitrary and can be chosen to reabsorb the term $\frac{1}{4}\log(4\pi)$. 
The extra dependence in (\ref{zio}) on the K\"ahler modulus $t_2$ is indeed separated in an overall additional 
term which decouples from the one-point amplitudes
$\partial_\sigma\F$.
Using this regularization scheme, that is deleting the tadpole term and, in case of vanishing Wilson lines, regulating the 
corresponding divergent series, we finally have:
\[
\F_{cyl} = - \sum_{i \neq j = 1}^{N}\Theta(|u_{i,j}|^{2})\frac{1}{2}log\left|e^{\pi i (\theta_{i,j})^{2}\sigma}\theta_{1}\left(u_{i,j}|\sigma \right)\eta\left( \sigma\right)^{-1}\right| - 
\]
\begin{equation}\label{j}
- \left( N + \sum_{i \neq j = 1}^{N}(1 - \Theta(|u_{i,j}|^{2}))\right) \frac{1}{2}log\left(\frac{\sqrt{\sigma_{2} t_{2}}}{\sqrt{2}}\left|\eta(\sigma)\right|^{2} \right) 
\end{equation}
\[
\F_{m\ddot{o}b} = + \sum_{i = 1}^{N}\Theta(|u_{i}|^{2})\frac{1}{2}log\left|e^{\pi i (2\theta_{i})^{2}\sigma}\theta_{1}\left(2u_{i}|\sigma \right)\eta\left( \sigma\right)^{-1}\right| + 
\]
\begin{equation}\label{k}
+ \left( \sum_{i = 1}^{N}(1 - \Theta(|u_{i}|^{2}))\right) \frac{1}{2}log\left(\frac{\sqrt{\sigma_{2} t_{2}}}{\sqrt{2}}\left|\eta(\sigma)\right|^{2} \right) 
\end{equation}
\begin{equation}\label{l}
\F_{kle} = -\frac{1}{2}log\left(\sqrt{\sigma_{2} t_{2}}\left|\eta(\sigma)\right|^{2} \right) 
\end{equation}
where $\Theta(x)$ is the step function, zero for $x\leq 0$ and one for $x>0$.

Let us now make a couple of observations on the above results.
First, notice that all the above free energies satisfy at generic values of the Wilson line a standard
holomorphic anomaly equation in the form
$
\partial_\sigma\partial_{\bar\sigma}\F \sim \frac{1}{(\sigma-\bar\sigma)^2}
$
with a proportionality constant counting the number of states in the appropriate vacuum bundle.
In particular, at vanishing Wilson lines, we recover the results stated in \cite{evidence}.
A more accurate discussion on the holomorphic anomaly equation for general target spaces is deferred to the next section.

The second comment  concerns the interpretation of the amplitudes we just calculated in terms of the analytic
Ray-Singer torsion \cite{RS}.

This is defined as \cite{pestunwitten}
\begin{equation}
log T(V)=\frac{1}{2}\sum_{q=0}^d (-1)^{q+1}q log ~det' \Delta_{V\otimes\Lambda^q T^*X}
\label{torsion}
\end{equation}
where $d=dim_{\C}X$.
On an elliptic curve with complex structure $\sigma$ the analytic torsion of a flat line bundle ${\cal L}$ 
with constant connection $u$ is
given by (as it can be found in Theorem 4.1 in \cite{RS})
\bea
\T\left({\cal L}\right)= 
\begin{cases} 
\quad \left\vert e^{-\pi \frac{\left(Im u\right)^2}{Im \sigma}}
\frac{\theta_1(u|\sigma)}{\eta(\sigma)}\right\vert                                            & \mbox{if } u\not=0 \\
\quad                                                                                              & \quad \\ 
\quad \sqrt{Im\sigma}\vert\eta(\sigma)\vert^2                                                     & \mbox{if } u=0 
\end{cases}
\eea
where the second case corresponds to the trivial line bundle ${\cal O}$.

On an elliptic curve equipped with a flat vector bundle $E=\oplus_i {\cal L}_i$, 
an extension of the formula for the Ray-Singer torsion 
implies that
\begin{eqnarray}
\F_{cyl}&=&-\frac{1}{2}\sum_{i,j}log \T\left({\cal L}_i\otimes{\cal L}^*_j\right) =
-\frac{1}{2}log \T\left(E\otimes E^*\right)
\nonumber\\
\F_{m\ddot{o}b}&=&+\frac{1}{2}\sum_{i}log \T\left({\cal L}_i^2\right)=
+\frac{1}{2}log \T\left(diag(E\otimes E)\right)
\nonumber\\ 
\F_{kle}&=&-\frac{1}{2}log \T\left({\cal O}\right).
\label{rst2}
\end{eqnarray}
The possibility to rewrite one-loop topological string amplitudes for the  B-model in terms of the analytic Ray-Singer torsion 
on the target space also for the unoriented open sector will be discussed in more detail in section \ref{storsion}.

Let us notice that the chamber structure in the amplitudes (\ref{rst2}) reflects exactly the multiplicative properties 
of the Ray-Singer torsion under vector bundle sums $log \T(V_1\oplus V_2)=log \T(V_1)+log \T(V_2)$ in the specific case of 
the torus.
In fact, the limit of vanishing Wilson line corresponds to the gauge bundle $E=E'\oplus {\cal O}$ and therefore one 
finds $log \T(E' \oplus {\cal O})=log \T(E')+log \T({\cal O})$.

\section{Unoriented topological string amplitudes at one loop}
\label{hae}

\subsection{Holomorphic anomaly equations}
\label{hanom}

In last section we considered a B-model topological string on a torus. Now we will generalize the computation to a generic Calabi-Yau 
3-fold. Namely, we will follow the standard BCOV's computation \cite{bcov} to derive holomorphic anomaly equations for the amplitudes 
of the cylinder, the M\"obius strip, and the Klein bottle. 

Firstly, we compute the cylinder amplitude $\F_{cyl}$. We fix the conformal Killing symmetry and the A- or B-twist on the cylinder by inserting a derivative with respect to the right moduli, that is, K\"ahler moduli for A-model and complex structure moduli for B-model 
of Calabi-Yau moduli space. We get the anomaly equation for the unoriented string amplitude
\begin{equation}
 \frac{\partial}{\partial \bar t^{\bar i}}\frac{\partial}{\partial t^{j}}\F_{cyl}=\frac14\int^\infty_0 ds
\left\langle\int d^2z \left\{(G^++\overline G^-),\bar\phi^{[1]}_{\bar i}\right\}\int_l  (G^-+\overline G^+)\int_{l'}\phi_j^{(1)}\right\rangle,
\end{equation}
where 
\begin{equation}
  \label{eq:BRST}
  Q_{BRST}=G^++\overline G^-,
\end{equation}
and
\begin{equation}
  \bar\phi_{\bar i}^{[1]}:=\frac12[(G^+-\overline G^-) ,\bar\phi_{\bar i}], \quad\quad \phi_j^{(1)}:=\frac12[(G^--\overline G^+),\phi_j].
\end{equation}
The degeneration gives rise to two contributions --- open channel and closed channel. The open channel is 
\begin{equation}
  \label{eq:open_channel}
 \frac14\bar\partial_{\bar i}\partial_j {\Tr}_{open}(-1)^Flog ~ g_{tt^*},
\end{equation}
where the trace is taken on the open string ground states, $g_{tt^*}$ is the $tt^*$ metric for the open string. 

\begin{figure}[h]
  \centering
  \subfigure[disk two-point function]{ \label{fig:disc_two_point_solo}
  \input{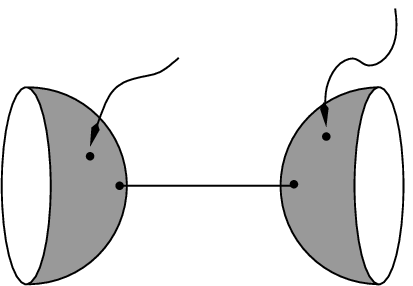}
                   }
\hspace{0.05\textwidth}
  \subfigure[tadpole for the cylinder degeneration]{ \label{fig:cylinder_tadpole_solo}
 \input{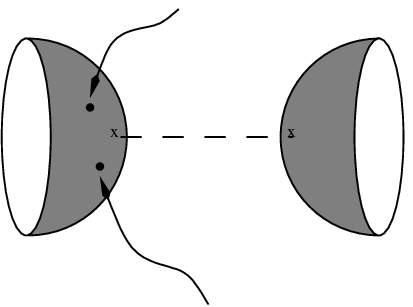}
                   }
\caption[cylinder_tadpole]{the degeneration of a cylinder}
\label{fig:cylinder_solo}
\end{figure}

For the closed channel, there are two cases.

i) The two operator insertions $\bar\phi_{\bar i}^{[1]}$ and $\phi_j^{(1)}$ are on different sides. It contributes to the equation by
\begin{equation}
  \label{eq:closed_twopoint}
  -\overline{\cal D}_{\bar i\bar k}{\cal D}_{jk}g^{\bar kk},
\end{equation}
where $g^{i\bar j}$ is the $tt^*$ metric for the closed string and ${\cal D}_{ij}$ is the disk two-point function (figure \ref{fig:disc_two_point_solo}).

ii) The two operator insertions are on the same side. It is a tadpole multiplied by a disk three-point function, where one operator insertion belongs to the wrong moduli, namely, complex structure moduli in A-model and K\"ahler moduli in B-model. In figure \ref{fig:cylinder_tadpole_solo}, we denote them as  $a$ and $\bar b$, the metric in between is $g^{a\bar b}$.

Next let us consider the amplitude's for a M\"obius strip
\begin{equation}
  \F_{m\ddot{o}b}=\int^\infty_0\frac{ds}{4s}{\Tr}[\P (-1)^FFe^{-2\pi sH}],
\end{equation}
where $\P$ is the involution operator. 

\begin{figure}[h]
  \centering
  \includegraphics[width=0.2\textwidth]{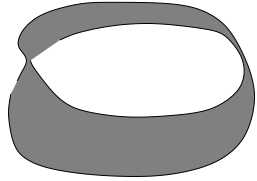}
  \caption{the pinching of a M\"obius strip}
  \label{fig:pinching_moebius}
\end{figure}

The holomorphic anomaly equation is then
\begin{equation}
\bar\partial_{\bar i}\partial_j \F_{m\ddot{o}b}=\frac14\int^\infty_0 ds\left\langle \P \int d^2z \left\{(G^++\overline G^-), \bar\phi^{[1]}_{\bar i}\right\}\int_l (G^-+\overline G^+)\int_{l'}\phi_j^{(1)} \right\rangle.
\end{equation}
Now the degeneracy has two types. One is the pinching of the strip, it gives rise to a contribution
\begin{equation}
  \label{eq:pinching_moebius}
  \frac14\bar\partial_{\bar i}\partial_j{\Tr}_{open}(-1)^F{\cal P}  log ~ g_{tt^*}.
\end{equation}
The only difference between the pinching of a cylinder and of a M\"obius strip (figure \ref{fig:pinching_moebius}), is the insertion 
of the involution operator $\P$ acting on the remaining strip amplitude. 

\begin{figure}[h]
  \centering
   \subfigure[crosscap two-point function and disk two-point function]{\label{fig:left_crosscap_solo}
   \input{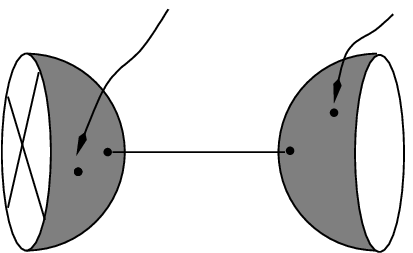}
             }
\hspace{0.1\textwidth}
  \subfigure[tadpole for the M\"obius degeneration]{\label{fig:left_crosscap_tadpole_solo}
  \input{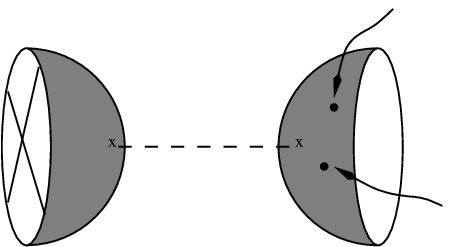}
             }
\caption{the degeneration of a M\"obius strip}
\label{fig:moebius_solo}
\end{figure}

The remaining degeneration amounts to remove the boundary from the M\"obius strip. There are two cases.

i) The two operator insertions are on the different sides (figure \ref{fig:left_crosscap_solo}). It gives rise to a disk two-point function multiplied by a crosscap two-point function
\begin{equation}
  \label{eq:disk_crosscap_twopoint}
  -(\overline {\cal C}_{\bar i\bar k}{\cal D}_{jk}+{\cal C}_{jk}\overline {\cal D}_{\bar i\bar k}) g^{k\bar k}.
\end{equation}
ii) The two operator insertions are on the same side (figure \ref{fig:left_crosscap_tadpole_solo}). It is a tadpole multiplied by a crosscap three-point function or a crosscap tadpole multiplied by a disk three-point function with one wrong modulus.

\begin{figure}[h]
  \centering
  \subfigure[two crosscap two-point functions]{\label{fig:klein_twopoint}
  \input{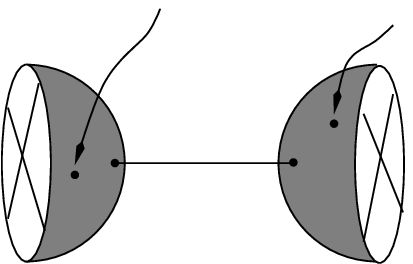}
            }
\hspace{0.1\textwidth}
  \subfigure[tadpole for the Klein degeneration]{\label{fig:kleintadpole}
  \input{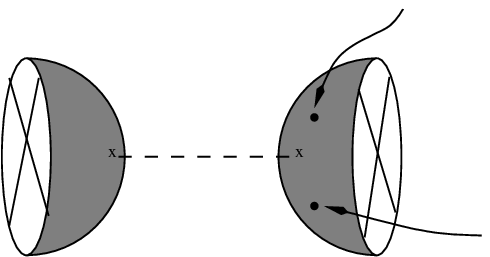}
            } 
  \caption{one degeneration of a Klein bottle}
  \label{fig:klein_solo}
\end{figure}

Finally, for the Klein bottle we have
\begin{equation}
  \F_{kle}=\int^\infty_0\frac{ds}{4s}{\Tr}[\P (-1)^F F e^{-2\pi sH}].
\end{equation}
There are two degenerations. Firstly, we consider the degeneration that splits the Klein bottle to two crosscaps. Again we have two cases.

i) The two operator insertions are on different sides (figure \ref{fig:klein_twopoint}).  It gives rise to two crosscap two-point functions
\begin{equation}
  \label{eq:klein_twopoint}
  -{\cal C}_{ik}\overline {\cal C}_{\bar j\bar k}g^{k\bar k}
\end{equation}

ii) The two operator insertions are on the same side (figure \ref{fig:kleintadpole}). It gives rise to  a crosscap tadpole multiplied by a crosscap three-point function with one wrong operator insertion.

Secondly, let us consider the complex double of the Klein bottle. Since this is a torus, the holomorphic anomaly equation is inherited from the torus. The only difference is that instead of a Yukawa coupling, we obtain an involution operator acting on the chiral/twisted chiral rings. The doubling torus degeneration gives rise, keeping into account a further factor $1/2$ from left/right projection, to
\begin{equation}
  \label{eq:Klein}
  \frac18 {\Tr}_{closed} [\P\overline C_{\bar i} C_j].
\end{equation}
This term corresponds to 
\begin{equation}
  \label{eq:quillen_klein}
 \frac18 \bar\partial_{\bar i}\partial_j {\Tr}_{closed}\P log ~ g,
\end{equation}
where $g$ is the $tt^*$ metric for the closed string.
\subsection{The derivative of the string amplitudes with respect to the wrong moduli}

In previous subsection we discussed about the anti-holomorphic dependence of one-loop open string amplitudes of the right moduli $\bar t^{\bar i}$. 
We can also calculate the derivative $\partial_i \F$ with respect to the wrong moduli $y^p$'s. 
Now we will study the different amplitudes separately. 

Firstly, we can consider what is the wrong moduli dependence of $\partial_i\F_{cyl}$.
\begin{equation}
\frac{\partial}{\partial t^{i}}\frac{\partial}{\partial y^{ p }}\F_{cyl}=\frac14\int^\infty_0 ds
\left\langle\int_l  (G^-+\overline G^+)\int_{l'} \phi^{(1)}_{i}\int d^2z\left\{(G^++\overline G^-),  
[G^-,\varphi_{p}]\right\}\right\rangle,
\end{equation}
where we use the same notation $\phi_{i}^{(1)}=\frac12[(G^--\overline G^+), \phi_{i}]$. We can check that this operator carries charge $1$.  
We define $\varphi^{(1)}_p=[G^-, \varphi_p]$, which has charge $-1$. 
Then we will perform a similar analysis as in the previous subsection. 

1) For the degeneration as the pinching of the two boundaries, we obtain
\begin{eqnarray}
  \label{eq:wrong_pinching}
&& \eta^{\alpha\beta}\left.\left\langle {\cal O}_\alpha(-\infty) \int_{l'}\phi^{(1)}_i\int d^2z\varphi^{(1)}_p (z)
{\cal O}_\beta(+\infty)\right\rangle\right|_{s\rightarrow\infty}\\
&=&\left.\frac12\eta^{\alpha\beta}\left[\langle\alpha |\int_{l'}(G^--\overline G^+)
  \phi_i\int d^2z[G^-,\varphi_p]|\beta\rangle-\langle\alpha |\int_{l'}
  \phi_i (G^--\overline G^+)\int d^2z[G^-,\varphi_p]|\beta\rangle \right]\right|_{s\rightarrow\infty}\nonumber
\end{eqnarray}where $\alpha, \beta$ are open string ground states and $\eta^{\alpha\beta}$ is the open string topological metric. 
This amplitude is independent of the time ($s$) position of the line $l'$, so we can put it in the center of the infinite strip. 
Thus the first piece of (\ref{eq:wrong_pinching})  is zero, because the $|\alpha\rangle$ state is projected to zero energy state 
by $e^{-2\pi sH}$ for $s\rightarrow\infty$, and so annihilated by $G^--\overline G^+$. 
The second piece is also zero, because now the $|\beta\rangle$ state is annihilated by $G^--\overline G^+$ for the same reason. 
Notice that the position of $[G^-, \varphi_p]$ does not matter, since it anti-commutes with $G^--\overline G^+$. 
Comparing with the right moduli case (\ref{eq:open_channel}),  we obtain 
\begin{equation}
  \label{eq:wrong_quillen}
  \partial_i\partial_p {\Tr}_{open}(-1)^Flog ~g_{tt^*}=0. 
\end{equation}

\begin{figure}[h]
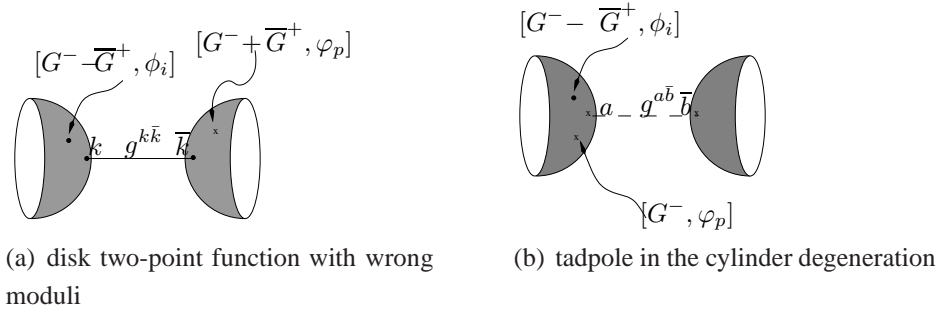

  \centering
  \subfigure[disk two-point function with wrong moduli]{\label{fig:disc_two_point}
  \input{disc_two_point.tex}
   }
\hspace{0.05\textwidth}
  \subfigure[tadpole in the cylinder degeneration]{\label{fig:cylinder_tadpole}  
  \input{cylinder_tadpole.tex}
  }
\label{fig:cylinder_wrong_moduli}
\caption{one degeneration of a cylinder}
\end{figure}

2) The second degeneration is the removing of a boundary from the cylinder. As before, there are two cases. 

i) The two operators insertions are on different sides (figure \ref{fig:disc_two_point}). 
Since $\phi_{i}^{(1)}$ and $\varphi_{p}^{(1)}$ have charges  $1$ and $-1$ respectively, 
in order to get charge $3$ or $-3$ on the disk we need to project the ground states to $(1, 1)$ and $(-1, -1)$ respectively. 
On one disk which has the wrong type of operator insertion $\varphi_p^{(1)}$, we can turn $\varphi_p^{(1)}$ into $[G^-+\overline G^+, \varphi_p]$. We know  that $G^-$ and $\overline G^+$ annihilate $(a, a)$ rings,  and $G^-+\overline G^+$ vanishes on the boundary. Therefore, this diagram does not contribute to the holomorphic anomaly equation. 

ii) The two operators insertions are on the same side (figure \ref{fig:cylinder_tadpole}). Again we obtain a tadpole multiplied 
by a disk three-point function. 

Secondly, we consider the M\"obius strip. It contains two cases: 1) The pinching of the boundary (see figure \ref{fig:pinching_moebius})
\begin{equation}
  \label{eq:pinching_moebius_wrong}
\eta^{\alpha\beta}\left.\left\langle  {\cal O}_\alpha(-\infty) \P \int_{l'}\phi^{(1)}_i\int d^2z\varphi^{(1)}_p (z)
{\cal O}_\beta(+\infty)\right\rangle\right|_{s\rightarrow\infty}.
\end{equation}
According to the similar argument as the case of the cylinder (\ref{eq:wrong_pinching}), we get zero.

\begin{figure}[h]
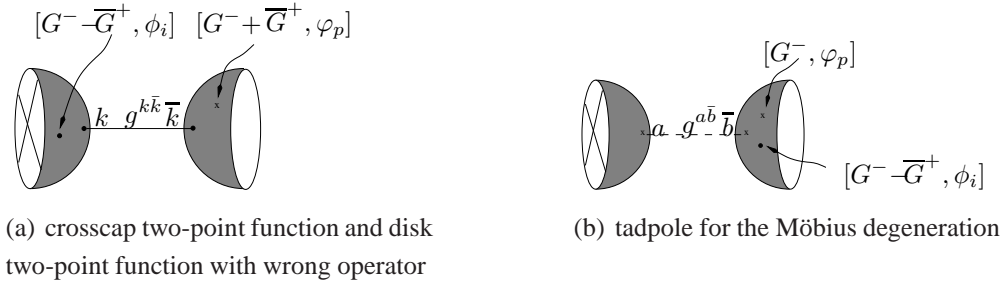

  \centering
  \subfigure[crosscap two-point function and disk two-point function with wrong operator]{\label{fig:left_crosscap}
  \input{left_crosscap.tex}
  }
\hspace{0.1\textwidth}
  \subfigure[tadpole for the M\"obius degeneration]{\label{fig:left_crosscap_tadpole}
  \input{left_crosscap_tadpole.tex}
  }
\label{fig:moebius_wrong_moduli}
\caption{one degeneration of a M\"obius strip}
\end{figure}

2) The removing of the boundary from the M\"obius strip.

i) One operator insertion is near the boundary, and the other is away from the boundary 
(figure \ref{fig:left_crosscap}).  If $\varphi_p^{(1)}$ is near the boundary, the degeneration for that disk will be a $(a, a)$ ring
inserted on the disk. From the same argument as for the cylinder, the disk two-point function is zero. If $\varphi_p^{(1)}$ is away 
from the boundary, namely, it is inserted on the crosscap, then that function is also zero. 

ii) The two operators insertions are on the same side (figure \ref{fig:left_crosscap_tadpole}). We obtain a tadpole multiplied by 
crosscap three-point insertions, or a crosscap multiplied by disk three-point insertions. 

\begin{figure}[h]
  \centering
  \input{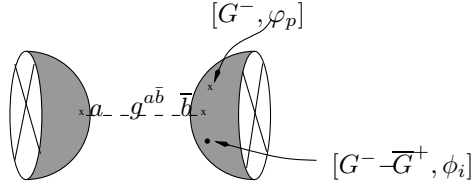}
\caption{one degeneration of a Klein bottle}
  \label{fig:klein_tadpole}
\end{figure}

Finally, for the Klein bottle, there is only one contribution. That is when the two insertions are on one side, we get the following 
diagram (figure \ref{fig:klein_tadpole}).

\subsection{Tadpole cancellation at one-loop}
\label{tadpole_oneloop}

\begin{figure}[h]
  \centering
  \includegraphics[width=0.4\textwidth]{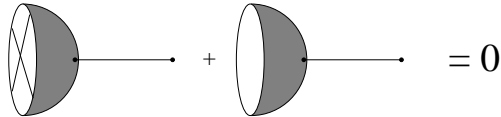}
  \caption{tadpole cancellation for one-loop}
  \label{fig:tadpole_cancellation}
\end{figure}

When we add up the holomorphic anomaly equations for the cylinder, M\"obius strip, and Klein bottle,
requiring tadpole cancellation 
(figure \ref{fig:tadpole_cancellation}), we get 
\begin{eqnarray}
  \label{eq:hol_anomaly}
  \bar\partial_{\bar i}\partial_j [\F_{cyl}+\F_{m\ddot{o}b}+\F_{kle}]&=&\frac18 \bar\partial_{\bar i}\partial_j {\Tr}_{closed} 
\left[\P log ~ g  \right]-\overline\Delta_{\bar i\bar k}\Delta_{jk}g^{k\bar k}, \nonumber\\
&&+\frac{1}{4}\bar\partial_{\bar i}\partial_j {\Tr}_{open} \left[(-1)^F (1+\P)log ~ g_{tt^*}\right]\\
  \partial_i\partial_p [\F_{cyl}+\F_{m\ddot{o}b}+\F_{kle}]&=&0, 
\end{eqnarray}where $\Delta_{ij}={\cal D}_{ij}+{\cal C}_{ij}$ is the sum of the disk and the crosscap two-point function. 

Eq.(\ref{eq:hol_anomaly}) reproduces the results stated in \cite{evidence} and extend
them to the presence of non-trivial open string moduli.  

\section{Unoriented one-loop amplitudes as analytic torsions}
\label{storsion}

In this section we discuss B-model unorientable one-loop amplitudes for generic Calabi-Yau threefolds
and provide a geometrical interpretation of them in terms of holomorphic torsions
of appropriate vector bundles.
 
Let us consider the Klein bottle amplitude first. As we have already seen,
this is given by the insertion of the involution operator ${\cal P}$ in the unoriented closed string trace
as
\begin{equation}
\label{klein0}
{\cal F}_{kle}= \int_0^\infty \frac{ds}{4s} {\Tr}_{{\cal H}_c} (-1)^F F {\cal P} e^{-2\pi sH_c}
\end{equation}
which we can compute as follows. We recall from \cite{bcov} that
the closed topological string Hilbert space is given by
$
{\cal H}_c
=
\Lambda^\bullet TX \otimes \Lambda^\bullet \bar T^*X=
\bigoplus_{p,q}\Lambda^p TX \otimes \Lambda^q \bar T^*X
$, where $TX$ and $\bar T^* X$ denote the holomorphic tangent bundle and anti-holomorphic
cotangent bundle respectively. 
At the level of the worldsheet superconformal field theory
these spaces are generated by the zero modes of the $\eta^{\bar I}$ and $\theta_{I}=\theta^{\bar I}g_{\bar I I}$
fermions respectively. The parity ${\cal P}$ acts as ${\cal P}\eta^{\bar I}=\eta^{\bar I}$ and 
${\cal P}\theta_{I}=-\theta_{I}$ \cite{horibrunner}. It is thus clear that the projection operator acts as ${\cal P}=(-1)^p$
on the closed string Hilbert space. 
By inserting in (\ref{klein0}) the expressions for the total fermion number $F=F_L+F_R=q-p$, a factor of $\frac{1}{2}$
which takes care of left/right identification 
and the closed string Hamiltonian in terms of the Laplacian $H=\Delta_{p,q}$, we get 
$$
{\cal F}_{kle}=\frac{1}{8}\sum_{p,q}(-1)^{q} q~ log\left( {\det}'\Delta_{p,q}\right)
=-\frac{1}{4}\sum_n log \T \left(\Lambda^n \bar T^*X\right)=
-\frac{1}{4}log \T\left(\Lambda^\bullet \bar T^*X\right)
$$
in terms of the analytic Ray-Singer torsion $\T(V)$ of the bundle $V=\Lambda^\bullet \bar T^*X$.

The cylinder amplitude is given by
$$
{\cal F}_{cyl}= \int_0^\infty \frac{ds}{4s} {\Tr}_{{\cal H}_o} (-1)^F F e^{-2\pi sH_o}
$$
where the assignment of the Chan-Paton factors 
selects ${\cal H}_o=\oplus_p\Lambda^p TX \otimes E\otimes E^*$ and the Hamiltonian 
$H_o=\Delta_{p,E\otimes E^*}$ is the corresponding Laplacian.
The result is (as already found in \cite{bcov})
$$
{\cal F}_{cyl}=-\frac{1}{2}log\T\left(E\otimes E^*\right).
$$

The last term to compute is the M\"obius strip amplitude that is
$$
{\cal F}_{m\ddot{o}b}= \int_0^\infty \frac{ds}{4s} {\Tr}_{{\cal H}_o} (-1)^F F {\cal P} e^{-2\pi sH_o}
$$
The only issue to discuss here is how to compute the trace with the ${\cal P}$ insertion. As explained in the appendix the trace over $\P$, for a $Sp(N/2)$ bundle, selects the $\diag(E\otimes E)$ states with $-1$ eigenvalue. 
This is the only non trivial action of the ${\cal P}$ operator 
on the Hilbert space. Indeed, the boundary conditions project away the $\theta_{\bar I}$'s and we are left with 
the $\eta^I$'s only, on which ${\cal P}$ acts as the identity.
Thus we get 
$$
{\cal F}_{m\ddot{o}b}=+\frac{1}{2}log\T\left({\rm diag}(E\otimes E\right)).
$$

Notice that the above conclusions agree with the explicit calculations of section \ref{tad1}, once restricted to 
the $T^2$ target space.

\subsection{Wrong moduli independence and anomaly cancellation}
\label{torsion_bismut}

In this section we show that the decoupling of wrong moduli in the unoriented 
open topological string on a Calabi-Yau threefold $X$ is equivalent to the usual D-brane/O-planes
anomaly cancellation. This is performed for the B-model with a system of $N$ spacefilling D-branes.
These are described by
a Chan-Paton gauge bundle $E$ over $X$ with structure group $U(N)$. 
As it is well known however, in order to implement the orientifold projection, $E\sim E^*$
has to be real therefore reducing the structure group to $SO(N)$ or $Sp(N/2)$ 
if the fundamental representation is real or pseudo-real respectively.

Let us now calculate the variation of the unoriented topological string free energy at one loop
under variations of the K\"ahler moduli.
In order to do it, we use the Bismut formula \cite{bismut} for the variation of the Ray-Singer torsion 
under a change of the base and fiber metrics $(g,h)\to (g+\delta g, h+\delta h)$
\begin{equation}
\left.\frac{1}{2\pi}\frac{\partial}{\partial t}\right\vert_{t=0} log \T(V)
=
\left.\frac{1}{2}\int_X \frac{\partial}{\partial t}\right\vert_{t=0}
\left[
Td\left(\frac{1}{2\pi}\left(iR + t g^{-1}\delta g\right)\right)
Ch\left(\frac{1}{2\pi}\left(iF + t h^{-1}\delta h\right)\right)
\right]_{8}
\label{bismut}
\end{equation}

By throwing the Bismut formula against the whole unoriented string free energy
${\cal F}^u_{\chi=0}=
{\cal F}_{cyl}+{\cal F}_{m\ddot{o}b}+
{\cal F}_{kle}$
and specializing to the variations of the K\"ahler form only (that is at a fixed metric on the Chan-Paton holomorphic vector bundle) 
we get\footnote{Here and in the following calculations we insert for convenience a formal parameter $n_o$ which keeps track
of the number of crosscaps. It will be eventually put to $1$.} 
\begin{equation}
\sim\int_X\left\{ \left[(Ch(E))^2 - n_o Ch({\rm diag}(E\otimes E))\right]
\frac{\partial}{\partial t}[ Td(TX)]_{t=0} + \frac{1}{2}n_o^2 
\frac{\partial}{\partial t} \left[Td(TX)
Ch(\Lambda^\bullet T^*)\right]_{t=0}\right\}
\label{variation}
\end{equation}
We will use $ch_k(2E)=2^kch_k(E)$ and $ch_k(E^*)=(-1)^kch_k(E)$, so that for $E=E^*$, $ch_k(E)=0$ for $k$ odd. 
From the definitions\footnote{See the book \cite{hirzebruch} for the notation.}
\begin{equation}
Td(TX)=\prod_a \frac{\gamma_a}{1-e^{-\gamma_a}}
\end{equation}
\begin{equation}
Ch(\Lambda^\bullet T^*X)=\prod_a (1+e^{-\gamma_a})
\end{equation}
we rewrite $Td(TX)=e^{c_1(T)/2}\hat A(TX)$ and $Td(TX)Ch(\Lambda^\bullet T^*X)=2^3 L(TX)$.
Using the standard expansions 
\begin{eqnarray}
\hat A&=& 1 - \frac{2}{3}p_1 2^{-4} +\frac{2}{45}\left(-4p_2+7p_1^2\right) 2^{-8} + \ldots
\\ \nonumber
L&=& 1 + \frac{1}{3}p_1 2^{-2} +\frac{1}{45}\left(7p_2-p_1^2\right) 2^{-4} + \ldots
\label{expans}
\end{eqnarray}
in (\ref{variation}) we 
calculate the variations of the cohomology classes above and obtain
\begin{equation}
\delta {\cal F}^u_{\chi=0} = \left.\int_X \sum_{i=1}^2 C_i \frac{\partial}{\partial t}p_i\right|_{t=0}
\label{variation2}
\end{equation}
with 
\begin{eqnarray}
C_1&=&-\frac{2^{-3}}{3}J_4(E)+\frac{2^{-6}\cdot 7}{45}p_1J_0(E)-\frac{2^{-1}}{45}n_o^2p_1\label{coeff0}
\\
C_2&=&-\frac{2^{-5}}{45}J_0(E)+\frac{2^{-2}\cdot 7}{45}n_o^2
\label{coeff}
\end{eqnarray}
where $J(E)=(Ch(E))^2-n_oCh({\rm diag}(E\otimes E))=J_0(E)+J_4(E)+\ldots$.
One verifies that, setting $n_o=1$, the vanishing of the coefficients (\ref{coeff0}) and (\ref{coeff}) is realized by
\bea 
ch_0(E) =8
\\ \nonumber
ch_2(E)= \frac{1}{4} p_1
\label{queste}
\eea
that can be rewritten in the more familiar form
\be
\sqrt{\hat A(TX)}Ch(E)-2^3\sqrt{\hat L(TX)}=0
\label{tadpo}\ee
that is\footnote{We denoted $\hat L=\prod_i\frac{\gamma_i/4}{th\left(\gamma_i/4\right)}$.}
the tadpole/anomaly cancellation condition for a system of spacefilling D-branes/O-planes
on a Calabi-Yau threefold \cite{anomacanc}.

\subsection{Quillen formula and holomorphic anomaly}
\label{Quillen}

In this subsection we compute the holomorphic anomaly equations of Section \ref{hae} from the expressions 
of the free energies in terms of Ray-Singer analytic torsion.

In order to do this, we apply the Quillen formula for torsions 
\begin{equation}
\partial\bar\partial log[T(V)]= \partial\bar\partial \sum_p \frac{(-1)^{p+1}}{2} log[\det g^{(p)}_V] -
\pi i \int_X \left[Td(TX)Ch(V)\right]_{(4,4)}
\label{Q}
\end{equation}
 where $\det g^{(p)}_V$ is the volume element in the kernel of $\bar\partial_V$ on $\Lambda^p T^* X\otimes V$ 
and $V$ is the relevant vector bundle for each contribution (that is $V_{cyl}=E\otimes E$, etc. see the beginning of the section) comparing 
with the first and the last terms in the r.h.s. of formula
(\ref{eq:hol_anomaly}).

In the notation of the previous subsection we get, up to the $\partial\bar\partial$-volume terms and setting $n_o=1$
\begin{eqnarray}
\partial\bar\partial {\cal F}^u_{\chi=0}= 
-\frac{\pi i}{2} \int_X \left[Td(TX)J(E)+\frac{1}{2}Td(TX)Ch(\Lambda^\bullet T^*X)  \right]_{(4,4)} + \\
+\frac{1}{4}\partial\bar\partial\left[\sum_p(-1)^p
\left(log\left[\det g^{(p)}_{E\otimes E^*}\right]]-log\left[\det g^{(p)}_{diag(E\otimes E)}\right]
+\frac{1}{2}log\left[\det g^{(p)}_{\Lambda^\bullet T^*X}\right]
\right)\right] \nonumber 
\label{Q1}
\end{eqnarray}
which we can calculate using the expansions (\ref{expans}) for the vector bundle $E$
satisfying (\ref{queste}).
The first line of (\ref{Q1}) is
\begin{eqnarray}
\pi i \int_X \left(
\left(ch_2(E)\right)^2 + \hat A_4\cdot 12 ch_2(E)+ 7\cdot 8\hat A_8+4L_8
\right)=0.
\label{Q2}
\end{eqnarray}
and vanishes. This result means that, once tadpoles are canceled, 
the $\bar\Delta\Delta$-term in (\ref{eq:hol_anomaly}) vanishes 
{\it for spacefilling branes/orientifolds}.
This is in agreement with the result for $T^2$ target found in section \ref{tad1}.


\section{Tadpole cancellation at all loops}
\label{all}

\subsection{Compactification of the moduli space of Riemann surfaces with boundaries}
\label{comp}

\begin{figure}[h]
  \centering
  \includegraphics[width=0.5\textwidth]{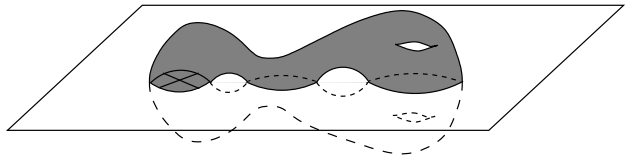}
  \caption{one example of an anti-holomorphic involution on $\Sigma$}
  \label{fig:antiholomorphic_involution}
\end{figure}

The moduli space of Klein surfaces with boundaries $\Sigma$ can be usefully described by referring 
to the notion of complex double $\left(\Sigma_{\mathbb C},\Omega\right)$, that 
is a compact orientable connected Riemann surface with an anti-holomorphic involution $\Omega$
(see figure \ref{fig:antiholomorphic_involution}).

The topological type of $\Sigma=\Sigma_{\mathbb C}/\Omega$ is classified by the fixed locus $\Sigma_{\mathbb R}$
of the involution \cite{natanzon}. 
If $\Sigma_{\mathbb R}=\emptyset$, then $\Sigma$ is non orientable and without boundaries, while if
$\Sigma_{\mathbb R}$ is not empty, then $\Sigma$ has boundaries. In the latter case, 
$\Sigma$ is orientable
if $\Sigma_{\mathbb C}\setminus\Sigma_{\mathbb R}$  is not connected and non orientable otherwise.

We recall that on a local chart $z\in{\complex}^*$, the anti-holomorphic involution acts as
$\Omega_\pm(z)=\pm\frac{1}{\bar z}$. The involution $\Omega_+$
has a non empty fixed set with the topology 
of a circle, which after the quotient becomes a boundary component. 
The involution $\Omega_-$ doesn't admit any fixed point and leads to a crosscap.

\begin{figure}[h]
  \centering
  \includegraphics[width=0.8\textwidth]{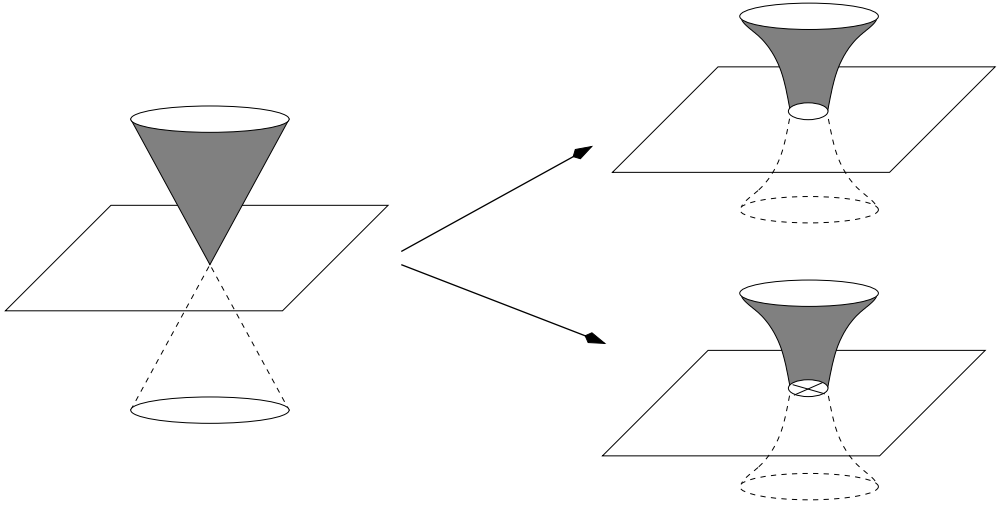}
  \caption{the complete resolution of a cone singularity in the doubling space}
  \label{fig:cone}
\end{figure}

The compactification of the moduli space of open Klein surfaces can be studied from the point of 
view of the complex double \cite{KL}. 
In this context, the boundary is given as usual by nodal curves, but with respect to the closed orientable 
case there are new features appearing due to the quotient.
In particular, nodes belonging to $\Sigma_{\mathbb R}$ can be smoothed either as boundaries or as crosscaps
(see figure \ref{fig:cone}).
Thus the moduli spaces of oriented and non orientable surfaces intersect at these boundary components
of {\it complex} codimension one.

\begin{figure}[h]
  \centering
  \includegraphics[width=0.8\textwidth]{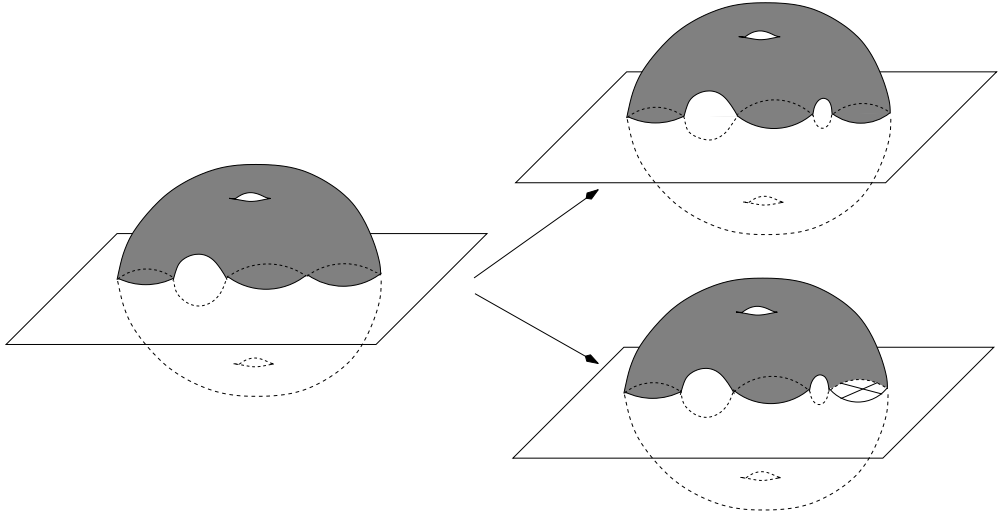}
  \caption{the resolution of a pinching point in the doubling space}
  \label{fig:pinching_doubling}
\end{figure}

Actually there are also boundary components of {\it real} codimension one
which are obtained when the degenerating 1-cycle of the complex double intersects 
$\Sigma_{\mathbb R}$ at points.
In this cases, one obtains the boundary open string degenerations as described in \cite{haom}.
The resolution of the real boundary nodes can be performed either as straight strips 
or as twisted ones. For example, when we have  colliding boundaries, their singularity can be resolved
either as splitting in two boundary components or as splitting in a single boundary and a crosscap
(see figure \ref{fig:pinching_doubling}). 
Thus the moduli space of oriented and unoriented surfaces intersect also along these components.
For a more detailed and systematic description, see \cite{chiuchu}.

More precisely, as discussed also in \cite{blau}, the moduli space of the quotient surface $\Sigma$ is obtained by 
considering the relative Teichm\"uller space $T(\Sigma_\complex,\Omega)$, that is the $\Omega$-invariant locus of 
$T(\Sigma_\complex)$, modding the large diffeomorphisms $\Gamma(\Sigma_\complex,\Omega)$ 
which commute with the involution $\Omega$ 
\be
{\cal M}_\Sigma=T(\Sigma_\complex,\Omega)/\Gamma(\Sigma_\complex,\Omega)
\label{moduli}\ee

\begin{figure}[h]
  \centering
  \psfrag{tau}{$\tau$}
 \includegraphics[width=0.3\textwidth]{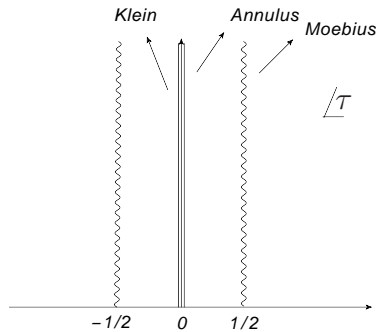}
  \caption{different involutions represent in the moduli space of Torus}
  \label{fig:involution_F1}
\end{figure}

Let us consider as an example the case of null Euler characteristic.
In this case the complex double is a torus and the annulus, M\"obius strip and Klein bottle
can be obtained by quotienting different anti-holomorphic involutions. 
The conformal families of tori admitting such involutions are 
Lagrangian submanifolds in the Teichm\"uller space of the covering torus modded by\footnote{The other generator $S$ of 
$\Gamma(T^2)=PSL(2,\mathbb{Z})$ is not quotiented because it does not commute with the involutions.} 
the translations $\tau\to\tau+1$
$\left\{\tau\in\complex | Im(\tau)>0, -\frac{1}{2}\leq Re(\tau)\leq \frac{1}{2}\right\}$.
These are vertical straight lines at $Re(\tau)=0$ for the annulus and the Klein bottle while
at $Re(\tau)=\pm\frac{1}{2}$ for the M\"obius strip
\footnote{Notice that the annulus and the Klein bottle are distinguished by different anti-holomorphic involutions.}
(see figure \ref{fig:involution_F1}).
Notice that all vertical lines meet at $\tau=i\infty$ which is the intersection point of the different moduli spaces.

At a more general level, one should similarly discuss the moduli space of holomorphic 
maps from the worldsheet $\Sigma$ to the Calabi-Yau space $X$ with involution $\sigma$ in terms of 
equivariant maps $(\Sigma_{\mathbb C},\Omega) \to (X, \sigma)$ \cite{KL}. The above discussion suggests that 
the proper definition of open topological strings can be obtained by summing over all possible inequivalent 
involutions of $\Sigma_{\mathbb C}$.  In particular one should include the contribution of non-orientable surfaces
in order to have a natural definition of the compactification of the space of stable maps.
 Actually, once the perturbative expansion of the string amplitudes
is set in terms of the Euler characteristic of the worldsheet, we have to
sum over all possible contributions ${\cal F}_\chi=\sum_{g,h,c|\chi=2-2g-h-c}{\cal F}_{g,h,c}$
at given genus $g\geq 0$, boundary number $h\geq 0$ and crosscaps number $0\leq c\leq 2$.

At fixed Euler characteristic, the set of Riemann surfaces admitting an anti-holomorphic involution
is a Lagrangian submanifold $L_\Omega$ of the Teichm\"uller space of the complex double $T_{\Sigma_{\mathbb C}}$
as in formula (\ref{moduli}).
Actually it might happen that the same Lagrangian submanifold corresponds to Riemann surfaces admitting
inequivalent involutions which have to be counted independently, as for the example of the annulus and
Klein bottle that we just discussed.
The complete amplitude is then given schematically as
$$
{\cal F}^u_\chi= \sum_\Omega
\int_{T_{\Sigma_{\mathbb C}}} \delta(L_\Omega)
\int_{\{\phi:\Sigma\to X, \phi\circ\Omega=\sigma\circ\phi\}}
|\mu G^-|^{3\chi}
$$
which provides a path integral representation for open/unoriented topological string amplitudes.

The above is the counterpart in topological string of the well-known fact that in open superstring theory 
unoriented sectors are crucial in order to obtain a consistent (i.e. tadpole and anomaly free) theory at all loops
\cite{BS}.
Evidence of these requirements has been found from a computational point of view in \cite{evidence}
where the contribution of unoriented surfaces has been observed to be necessary to obtain integer
BPS counting formulas for A-model open invariants on some explicit examples.
Let us remark that this picture applies to any compact or non compact Calabi-Yau threefold in principle. 
It might happen, however, that in the non compact case for some specific D-brane geometries tadpole cancellation 
can be ensured by choosing suitable boundary conditions at infinity 
so that the orientable theory is consistent by itself as in the case of \cite{KL,SV}.

\subsection{Local tadpole cancellation and holomorphic anomaly}
\label{local}


As we have seen in the Section \ref{comp}, non-orientable Riemann surfaces should be included in 
order to provide a consistent compactification of the moduli space of open strings.
It was found in \cite{jie} that a dependence on wrong moduli appears when one considers
holomorphic anomaly equations for orientable Riemann surfaces with boundaries.
However, it follows from the discussion of Section \ref{comp} that whenever we consider a closed string degeneration
in which one of the boundaries shrinks, there is always a corresponding component in the boundary
of the complete moduli space where one crosscap is sent to infinity.  
Therefore, we always have this type of degeneration 
\begin{equation}
  \label{eq:tadpole_crosscap}
  {\cal A}^a(\langle\omega_a|B\rangle +\langle\omega_a|C\rangle),
\end{equation}
where $|B\rangle$ and $|C\rangle$ are the boundary and crosscap state respectively, 
$\omega_a$ is the operator inserted in the degenerated point which corresponds to a wrong modulus, 
${\cal A}^a$ is the amplitude of the remaining Riemann surface with a wrong moduli operator insertion.  
Tadpole cancellation implies
\begin{equation}
  \label{eq:tadpole_cancellation}
  \langle\omega_a|B\rangle +\langle\omega_a|C\rangle=0.
\end{equation}
which ensures the cancellation of the anomaly of \cite{jie} at all genera.
This cancellation has a simple geometrical interpretation in the A-model:
in this case, we can have D6-branes and O6-planes wrapping 
3-cycles of the Calabi-Yau 3-fold $X$, and the condition (\ref{eq:tadpole_cancellation}) reads 
\begin{equation}
  \label{eq:tadpole_geometry}
  \langle \omega_a|B\rangle+\langle\omega_a|C\rangle=\partial_{y^a}\left(\int_L \Omega^{(3,0)}+ \int_{X^\sigma}\Omega^{(3,0)}\right)=0,
\end{equation}where $L$ is a Lagrangian 3-cycle, $\Omega^{(3,0)}$ is the holomorphic 3-form, and $X^\sigma$ is the fixed point set of the involution $\sigma: X\rightarrow X$. From 
(\ref{eq:tadpole_geometry}) we can interpret the local cancellation of the wrong moduli dependence (\ref{eq:tadpole_crosscap})
as a stability condition for the vacuum against wrong moduli deformations.

\section{Conclusions}
\label{conclusion}

In this paper we discussed the issue of tadpole cancellation in the context of unoriented
topological strings, and showed from Super Conformal Field Theory arguments that this corresponds
to the decoupling of wrong moduli at all loops.
We also provided a geometrical interpretation for unoriented B-model amplitudes at one loop in terms of analytic 
torsions of vector bundles over the target space.

Let us remark that the topological open A-model free energy is expected to provide a generating
function for open Gromov-Witten invariants. However, these have not been
defined rigorously yet, except for some particular cases \cite{KL,GZ,panda}.
We observe that the inclusion of unoriented worldsheet geometries turns out
to be natural also from a purely mathematical viewpoint.
In fact the compactified moduli spaces of open Riemann and Klein surfaces
have common boundary components (see Section \ref{comp}).
Thus string theory suggests that a proper mathematical definition
of open Gromov-Witten invariants should be obtained by including
non-orientable domains for the maps. Therefore one should consider
equivariant Gromov-Witten theory and sum over all possible involutions
of the complex double, up to equivalences.

There are several interesting directions to be further investigated,
the most natural being the study of holomorphic anomaly equations in presence of non-trivial
open string moduli. This can be obtained by extending the holomorphic anomaly equations
studied in \cite{haom} in order to include the contribution of non-orientable worldsheets.

Actually, our method is applicable only to cases in which the D-brane/orientifold set is 
modeled on the fixed locus of a target space involution. It would be quite interesting to be able to 
generalize it to a more general framework, that is to remove the reference to a given target space involution 
to perform the orientifold projection, in order to compare with 
some of the compact examples studied in \cite{j,Grimm,kw2,w5,joh}.

It would be also interesting to link our B-model torsion formulae to the A-model side where open strings on orientifolds have been
understood quite recently \cite{marino} to be the dual of coloured polynomials in the Chern-Simons theory.
This should also enter a coloured extension of the conjecture stated in \cite{df}.
Notice also that interpretation of open B-model one loop amplitudes in terms of analytic torsions could be extended to 
more general target space geometries. For example one could investigate whether the notion of twisted torsions introduced in
\cite{RW,MW} could provide a definition of B-model one loop amplitudes in the presence of $H$-fluxes and more in general with a target
of generalized complex type.
In such a context, our approach should lead to a generalization to open strings of the computation of exact gravitational threshold corrections
as in \cite{fhsv,hm}.

{\bf Acknowledgments}:
We thank A.~Brini, R.~Cavalieri, S.~Cecotti, H.-L.~Chang, J.~Evslin, H.~Liu, H.~Ooguri, J.~Walcher and J.-Y. Welschinger 
for discussions and exchange of opinions. We thank S.~Natanzon for providing a copy of \cite{natanzon}.

\section*{Appendix}
Here we want to discuss in more detail the effect of the Chan-Paton factors to the amplitudes. 
The notation follows from \cite{Polchinski_I}. An open string state is generalized carrying two indices at the two ends, 
each one running on the integers from $1$ to $N$. This additional state is indicated as $\left|i,j\right\rangle$ $ (i,j = 1 ... N)$.

The worldsheet parity $\Omega$ is defined to act exchanging $i$ with $j$ and rotating them with a $U(N)$ transformation $\gamma$. 
This rotation is added simply because it is still a symmetry for the amplitudes. Thus we have
\begin{equation} \label{aa}
\Omega\left|i,j\right\rangle \equiv \gamma_{j\:l}\left|l,k\right\rangle\gamma^{-1}_{k\:i}
\end{equation}
Asking $\Omega^{2} = 1$ \cite{Polchinski_I} means requiring
\begin{equation} \label{ab}
\gamma^{T} = \pm \gamma
\end{equation}
Now if we do a base change of the kind
$\left|i,j\right\rangle \rightarrow \left|i',j'\right\rangle = U^{-1}_{i'\:k}\left|k,l\right\rangle U_{l\:j'}$
it transforms $\gamma$ in the new primed base so that
$\gamma \rightarrow U^{T}\gamma U$.
In particular choosing an appropriate $\left|i',j'\right\rangle $ base one can always transform $\gamma$ so that
\begin{equation} \label{ae}
\gamma = 1 \:\:\:\:\: or \:\:\:\:\: \gamma = \left( \begin{array}{cc} 0 & i \\ -i & 0 \end{array}\right)
\end{equation}
respectively in the $+$ or $-$ case of (\ref{ab}).
We start from the first case. There we can create the new base
$\left|a\right\rangle = \Lambda^{a}_{i\:j}\left|i,j\right\rangle$
using $N^{2}$ independent matrices, in our case the $N \times N$ real matrices.  Worldsheet parity action on the states 
$\left|a\right\rangle$ can be seen as an action on the coefficient $ \Lambda^{a}_{i\:j}$. Choosing them either symmetric or 
antisymmetric one has respectively $\frac{1}{2}(N^{2} + N)$ and  $\frac{1}{2}(N^{2} - N)$ of them. 
Since massless states transform with a minus under worldsheet parity, in order to create unoriented states one needs to couple these 
to Chan-Paton states $\left|a\right\rangle $ with antisymmetric coefficients. So a double minus gives a plus. Then the gauge field 
background, associated with those vertex operators, will be with values in the Lie Algebra of $N \times N$ anti-symmetric real matrices,
that is $SO(N)$.
From the spacetime effective action, with gauge field and matter in the adjoint, one has that the coupling of an  
$\left|a\right\rangle $ state with a generic  background $A = A_{b}\Lambda^{b}$ is of the kind
$\left[\Lambda^{a},A_{b}\Lambda^{b}\right]$.
If the background is diagonal with elements $a_{1} ... a_{N}$ and we consider the state $\left|a\right\rangle = \left|i,j\right\rangle$
this coupling gives an eigenvalue $+a_{i} - a_{j}$: the state $\left|i,j\right\rangle $  will shift the spacetime momenta as 
$p \rightarrow p + a_{i} - a_{j} $. This effect is  more precisely described changing the string action with the addition of a 
gauge field background, which will manifest itself inserting in the path integral a Wilson loop of the kind
\begin{equation} \label{af}
\prod_{k}{\Tr} e^{i\int_{\partial\Sigma_{k}}A_{\mu}\dot X^{\mu}}
\end{equation}
where the sum is other all the connected components of the boundary. If one creates an open string state with a vertex operator on a 
boundary with non trivial homology, the left $i$ Chan-Paton sweeps in space giving an Aharonov-Bohm phase (\ref{af}) $a_{i}$. 
The right $j$ Chan-Paton moves in the opposite direction on the same boundary and couples with a minus. For loop states with one 
Chan-Paton on a boundary and the second on another the situation is the same, always with one Chan-Paton moving along the orientation 
of the boundary and the other in the opposite \footnote{Notice that the state sweeping the loop should be consistent with the one 
created by a boundary vertex operator plus some string interaction}.

Now for any (constant) $SO(N)$  background one can always act with a rigid gauge transformation to put it in the form
\begin{equation} \label{ag}
\bigoplus_i a_i\left( \begin{array}{cc} 0 & 1 \\ -1 & 0
\end{array}\right)
\end{equation}
This is still $SO(N)$ so the worldsheet parity still acts simply exchanging the $i - j$ factors. If in addition one wants to diagonalize 
it one needs to act with a gauge transformation that will change the  $SO(N)$ form and so will have effects also on the shape of 
$\Omega$. In fact we can rewrite $\left|i,j\right\rangle =  U_{i\:k'}\left|k',l'\right\rangle U^{-1}_{l' \:j}$ so that
\[
\left|a\right\rangle =\Lambda^{a}_{i\:j}U_{i\:k'}\left|k',l'\right\rangle U^{-1}_{l' \:j} = U^{T}_{k'\:i}\Lambda^{a}_{i\:j}(U^{\dagger})^{T}_{j\:l'}\left|k',l'\right\rangle
\]
This in order to transform $\Lambda^{a} $ so to diagonalize our background. But, acting in this way, the base $\left|i,j\right\rangle$ 
has changed and then also the worldsheet parity (\ref{aa}) will be different. In particular
\[
\gamma ( = I )\rightarrow  U^{T}\gamma U ( = U^{T}U )
\]
When (\ref{ag}) is reduced to the simple two dimensional case the matrix $U^{T}$ which diagonalizes $A$ and $\gamma$ ( in the primed base ) are
\[
U^{T} =  \left( \begin{array}{cc} i/\sqrt{2} & 1/\sqrt{2} \\ -i/\sqrt{2} & 1/\sqrt{2} \end{array}\right) \:\:\:\:\:\: \gamma =  \left( \begin{array}{cc} 0 & 1 \\ 1 & 0 \end{array}\right)
\]
The computation of the cylinder is straightforward. We have to sum over all states, and different Chan-Paton indices will modify the 
Hamiltonian with the usual momentum shift. Instead if we want to compute the M\"obius strip  we should look for diagonal states of 
$\Omega $. It is easy to see that, in the $\left|i',j'\right\rangle $ base, they are $\left|1,2\right\rangle  $ and 
$\left|2,1\right\rangle$, both with eigenvalue $+ 1$\footnote{ In the $ \left|a\right\rangle $ base there are four, one with 
eigenvalue $- 1$, that is the diagonalized background itself, and three with $+1$. }. Each diagonal term in the trace will contribute 
both with its eigenvalue and with its own Hamiltonian. In our case the two states will change the momenta respectively as 
$p \rightarrow p + a_{1} - a_{2} $ and $p \rightarrow p + a_{2} - a_{1} $ where, for our diagonalized $SO(2)$  background, 
$a_{1} = ia$ and $a_{2} = -ia$. Generalization to higher $N$ is straightforward. Then we end up with our amplitudes. 

The $Sp(N/2)$ situation is even simpler. There the diagonal background is already an $Sp(N/2)$ algebra matrix if in the 
form\footnote{If one interchanges the positions of some diagonal elements, which can of course be done with a gauge transformation, 
that matrix is no longer $Sp(N/2)$.The cylinder is manifestly invariant under any such gauge transformation, but the M\"obius is not. 
In fact if one wants to compute the amplitude in the new background one should take care of the changing occurred to the worldsheet 
parity operator and find the new diagonal states with their gauge field couplings. Working properly the amplitude is of course 
invariant. } 
\[\diag\{a_1, \cdots, a_{N/2}, -a_1, \cdots, -a_{N/2}\}
\] 
 Therefore the worldsheet parity $\Omega$ is still the second of (\ref{ae}). Diagonal states are now $\left|i,i + N/2\right\rangle  $ or 
$\left|i + N/2,i\right\rangle  $ for $i = 1 ... N/2$, note both with negative eigenvalues. The contribution to the Hamiltonian is 
again $p \rightarrow p \pm 2a_{i} $.

\end{document}